\documentclass[
    ,final            
  ]
  {aipproc}

\layoutstyle{6x9}

\newcommand{\befig}{\begin{figure}}
\newcommand{\efig}{\end{figure}}
\newcommand{\betab}{\begin{table}}
\newcommand{\etab}{\end{table}}
\newcommand{\barray}{\begin{array}}
\newcommand{\earray}{\end{array}}
\newcommand{\be}{\begin{equation}}
\newcommand{\ee}{\end{equation}}
\newcommand{\bea}{\begin{eqnarray}}
\newcommand{\eea}{\end{eqnarray}}
\newcommand{\benn}{\begin{displaymath}}
\newcommand{\eenn}{\end{displaymath}}
\newcommand{\beann}{\begin{eqnarray*}}
\newcommand{\eeann}{\end{eqnarray*}}
\newcommand{\gtsim}{\gtrsim}
\newcommand{\ltsim}{\lesssim}
\newcommand{\lsim}{\buildrel<\over{_\sim}}
\newcommand{\gsim}{\buildrel>\over{_\sim}}
\newcommand{\gtrsim}{\gsim}
\newcommand{\lesssim}{\lsim}
\newcommand{\GeV}{\mathrm{GeV}}
\newcommand{\TeV}{\mathrm{TeV}}
\newcommand{\Mpc}{\mathrm{Mpc}}
\newcommand{\km}{\mathrm{km}}
\newcommand{\seconds}{\mathrm{s}}

\newcommand{\MPl}{\mathrm{M}_{\mathrm{P}}}
\newcommand{\gravitino}{{\widetilde{G}}}

\newcommand{\sel}{\ensuremath{\tilde{\mathrm{e}}}}

\newcommand{\smu}{\ensuremath{\tilde{\mu}}}
\newcommand{\stau}{{\widetilde \tau}}
\newcommand{\st}{\ensuremath{\tilde{\tau}}}
\newcommand{\quark}{\ensuremath{\mathrm{q}}}
\newcommand{\antiquark}{\ensuremath{\bar{\mathrm{q}}}}
\newcommand{\topquark}{\ensuremath{\mathrm{t}}}
\newcommand{\antitopquark}{\ensuremath{\bar{\mathrm{t}}}}

\newcommand{\gr}{\ensuremath{\tilde{G}}}
\newcommand{\Bi}{\ensuremath{\tilde{B}}}
\newcommand{\mgravitino}{m_{\widetilde{G}}}
\newcommand{\mgr}{m_{\widetilde{G}}}
\newcommand{\mst}{m_{\tilde{\tau}_1}}
\newcommand{\EM}{\mathrm{em}}
\newcommand{\HAD}{\mathrm{had}}

\newcommand{\NTP}{\mathrm{NTP}}
\newcommand{\TP}{\mathrm{TP}}

\newcommand{\CDM}{\mathrm{DM}}

\newcommand{\Reheating}{\mathrm{R}}
\newcommand{\Color}{\mathrm{c}}
\newcommand{\Weak}{\mathrm{L}}
\newcommand{\Hypercharge}{\mathrm{Y}}
\begin{document}

\title{Constraints on Gravitino Dark Matter Scenarios with Long-Lived
  Charged Sleptons}


\author{Frank Daniel Steffen}{
address={Max-Planck-Institut f{\"u}r Physik, 
F{\"o}hringer Ring 6,
D--80805 Munich, Germany}}

\begin{abstract}
  Considering scenarios in which the gravitino is the lightest
  supersymmetric particle and a charged slepton the next-to-lightest
  supersymmetric particle (NLSP),
  we discuss cosmological constraints on the masses of the gravitino
  and the NLSP slepton.
  The presented mass bounds are crucial for gravitino dark matter
  studies and potential gravitino signatures at future colliders.
\end{abstract}

\maketitle


The existence of the gravitino $\gravitino$---the gauge field of local
supersymmetry (SUSY) transformations---is an unavoidable implication
of SUSY theories containing gravity. As the spin-3/2 superpartner of
the graviton, the gravitino is an extremely weakly interacting
particle with couplings suppressed by inverse powers of the (reduced)
Planck scale $\MPl=2.4\times 10^{18}\,\GeV$. In the course of
spontaneous SUSY breaking, the gravitino acquires a mass $\mgravitino$
which depends on the SUSY breaking scheme.

We consider R-parity conserving scenarios in which the gravitino is
the lightest SUSY particle (LSP) and a charged slepton the
next-to-lightest SUSY particle (NLSP). In such scenarios, the
gravitino LSP is a promising dark matter
candidate~\cite{Borgani:1996ag,Bolz:1998ek,Asaka:2000zh,Bolz:2000fu,Feng:2004mt,Cerdeno:2005eu,Steffen:2006hw,Pradler:2006qh}.
Moreover, the NLSP typically has a long lifetime due to the extremely
weak couplings of the gravitino. Thus, the charged slepton NLSP can
provide striking signatures at future colliders which can lead to
evidence for gravitino dark
matter~\cite{Buchmuller:2004rq,Brandenburg:2005he,Steffen:2005cn,Martyn:2006as}.

In the following we discuss cosmological constraints on gravitino dark
matter scenarios with long-lived charged sleptons.  To be specific, we
focus on the case where the pure `right-handed' stau $\stau_{\mathrm
  R}$ is the NLSP. More detailed discussions can be found
in~\cite{Steffen:2006hw}.

Gravitino dark matter can originate from thermal production (TP) in
the very early Universe~\cite{Bolz:1998ek,Bolz:2000fu,Pradler:2006qh}
and from non-thermal production (NTP) in decays of the stau
NLSP~\cite{Borgani:1996ag,Asaka:2000zh,Feng:2004mt}. 
The gravitino density parameter is thus given by
$\Omega_{\gravitino}h^2=\Omega_{\gravitino}^{\TP}h^2+\Omega_{\gravitino}^{\NTP}h^2$
where
\bea
        \Omega_{\gravitino}^{\TP\hphantom{N}}h^2
        &=&
        \sum_{i=1}^{3}
        \omega_i\, g_i^2 
        \left(1+\frac{M_i^2}{3\mgr^2}\right)
        \ln\left(\frac{k_i}{g_i}\right)
        \left(\frac{\mgr}{100~\GeV}\right)
        \left(\frac{T_{\Reheating}}{10^{10}\,\GeV}\right)
        \ ,
\label{Eq:GravitinoDensityTP}
\\
        \Omega_{\gravitino}^{\NTP} h^2
        &=& 
        \mgravitino\, Y_{\st}\, s(T_0) h^2 / \rho_{\mathrm{c}}
        \ ,
\label{Eq:GravitinoDensityNTP}
\eea
with the Hubble constant $h$ in units of
$100~\km\,\Mpc^{-1}\seconds^{-1}$ and
$\rho_c/[s(T_0)h^2]=3.6\times 10^{-9}\,\GeV$.
In expression~(\ref{Eq:GravitinoDensityTP}), the gaugino mass
parameters $M_i=(M_1,M_2,M_3)$, the gauge couplings
$g_i=(g',g,g_\mathrm{s})$,
$\omega_i=(0.018,0.044,0.117)$, and $k_i=(1.266,1.312,1.271)$
are associated with the gauge groups U(1)$_\Hypercharge$,
SU(2)$_\Weak$, and SU(3)$_\Color$, respectively~\cite{Pradler:2006qh}.
Here $M_i$ and $g_i$ are understood to be evaluated at the scale given
by the reheating temperature after inflation $T_{\Reheating}$.
Expression~(\ref{Eq:GravitinoDensityNTP}) involves the stau NLSP yield
$Y_{\st}\equiv n_{\st}/s$, where $s$ is the total entropy density of
the Universe and $n_{\st}$ the number density that the stau NLSP would
have today, if it had not decayed. Assuming that the stau NLSPs freeze
out with a thermal abundance before their decay, one can work with the
simple relation~\cite{Asaka:2000zh,Steffen:2006hw}
\be
        Y_{\st}
        = 
        0.725 \times 10^{-12} 
        \left(\frac{m_{\st}}{1~\TeV}\right)
\label{Eq:YstauNoCo}
\ee
which is valid for a superparticle spectrum in which the stau NLSP
mass $m_{\st}$ is significantly below the masses of the lighter
selectron and the lighter smuon: $m_{\st} \ll m_{\sel_1,\smu_1}$.
Moreover, the lightest neutralino is assumed to be a pure bino with a
mass of $m_{\Bi}=1.1\,m_{\st}$. Coannihilation processes of sleptons
with binos are not taken into account.

The dark matter density~\cite{PDB2006}
$\Omega_{\CDM}h^2=0.105^{+0.014}_{-0.020}$ (95\%~CL) limits
$\Omega_{\gravitino}h^2$ from above. Because of the $T_{\Reheating}$
sensitivity of $\Omega_{\gravitino}^{\TP}$, the constraint
$\Omega_{\gravitino}\leq\Omega_{\CDM}$ implies upper limits on
$T_{\Reheating}$ that are crucial for our understanding of inflation
and baryogenesis~\cite{Cerdeno:2005eu,Steffen:2006hw}.

\begin{figure}
  \includegraphics[width=0.9\textwidth]{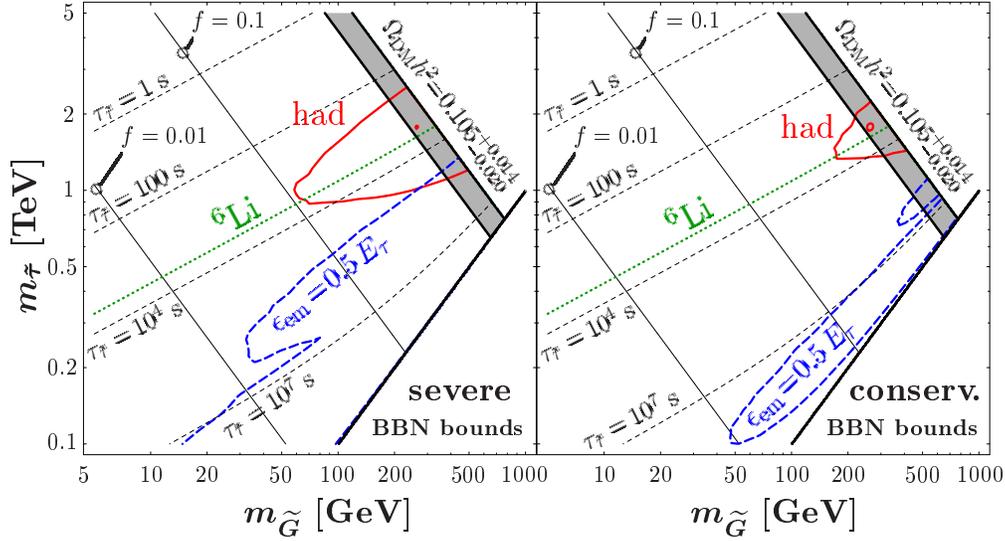}
\label{Fig:MassBoundsAHSNoCo}
\caption{Cosmological constraints on the masses of the gravitino LSP
  and the stau NLSP.}
\end{figure}

For given $\Omega_{\gravitino}^{\TP}$, the bound
$\Omega_{\gravitino}^{\NTP}\leq\Omega_{\CDM}-\Omega_{\gravitino}^{\TP}$
gives upper limits on $\mgr$ and $\mst$. The limits obtained
with~(\ref{Eq:YstauNoCo}) are shown in
Fig.~\ref{Fig:MassBoundsAHSNoCo}.
The grey band marks the region in which
$\Omega_{\gravitino}^{\NTP}\!\!\simeq\Omega_{\CDM}$.  The region above
the grey band is excluded because of $\Omega_{\gravitino} >
\Omega_{\CDM}$. Only $10\%$ ($1\%$) of $\Omega_{\CDM}$ is provided by
$\Omega_{\gravitino}^{\NTP}$ for scenarios that fall onto the thin
solid line labeled by $f=0.1$ ($f=0.01$). 
For these scenarios, $\Omega_{\gravitino}^{\TP}$ can give $90\%$
($99\%$) of $\Omega_{\CDM}$.

While thermally produced gravitinos typically have a negligible
free--streaming velocity today, gravitinos from late stau NLSP decays
can be warm dark matter for $m_{\st} \lesssim 5~\TeV$. Indeed, upper
limits on the free--streaming velocity from simulations and
observations of cosmic structures exclude
$m_{\st}\lesssim 0.7~\TeV$
for $\Omega_{\gravitino}^{\NTP}\!\!\simeq\Omega_{\CDM}$. However, such
scenarios (grey band) require anyhow $m_{\st}\gtrsim 0.7~\TeV$ and
could even resolve the small scale structure problems inherent to cold
dark matter; cf.~\cite{Steffen:2006hw} and references therein.

The stau NLSP lifetime $\tau_{\st}$ is governed by the 2-body decay
$\stau\to\tau\gravitino$,
\begin{equation}
        \tau_{\st} 
        \simeq \Gamma^{-1}(\stau\to\tau\gravitino)
        = \frac{48 \pi \mgr^2 \MPl^2}{m_{\st}^5} 
        \left(1-\frac{\mgr^2}{m_{\st}^2}\right)^{-4}
        \ ,
\label{Eq:StauLifetime}
\end{equation}
as obtained from the supergravity prediction of the partial width
$\Gamma(\stau\to\tau\gravitino)$ for $m_{\tau}\to 0$. In
Fig.~\ref{Fig:MassBoundsAHSNoCo} the thin dashed lines show
$\tau_{\st}=1$, $100$, $10^4$, and $10^7\seconds$ (from top to
bottom).

For stau NLSP decays occuring during or after big bang nucleosynthesis
(BBN), the Standard Model particles emitted in addition to the
gravitino can affect the abundances of the primordial light elements
(D, He, Li). Demanding that the successful BBN predictions are
preserved, upper bounds on electromagnetic and hadronic energy release
in decays of a species $X$,
$\xi_{\EM,\HAD} \equiv \epsilon_{\EM,\HAD}\, Y_{X}$,
have been obtained~\cite{Kawasaki:2004qu}; more conservative bounds on
$\xi_{\EM}$ can be found in~\cite{Cyburt:2002uv}. We use these limits
for $X=\stau_{\mathrm R}$ where $\epsilon_{\EM,\HAD}$ is the (average)
electromagnetic/hadronic energy emitted in a single $\stau_{\mathrm
  R}$ decay and $Y_{X}=Y_{\st}$.

The 2-body decay $\stau\to\tau\gravitino$ governs $\epsilon_{\EM}$.
For late decays, $\tau_{\st} \gtrsim 100~\seconds$, the emitted tau
decays before interacting electromagnetically. Since the neutrinos
from the $\tau$ decays are only weakly interacting, their effect on
the primordial nuclei is typically subleading. Accordingly,
$\epsilon_{\EM}=0.5 E_{\tau}$ is a reasonable estimate, where
$E_{\tau}=(m_{\stau}^2-\mgravitino^2+m_{\tau}^2)/(2\,m_{\stau})$
is the tau energy in the rest frame of the stau
NLSP~\cite{Feng:2004mt}. Moreover, the hadronic effects of the mesons
from the $\tau$ decays can be neglected for $\tau_{\st} \gtrsim
100~\seconds$ so that $\epsilon_{\HAD}(\stau\to\tau\gravitino)\simeq
0$.  Then, the 4-body decay of the stau NLSP into the gravitino, the
tau, and a quark--antiquark pair governs the constraints from late
hadronic energy injection~\cite{Steffen:2006hw}:
\be 
    \epsilon_{\HAD} (\stau_{\mathrm R}\to\tau\,\gravitino\,\quark\,\antiquark) 
        \equiv {1 \over \Gamma^{\rm{total}}_{\tilde{\tau}}} \,
        \int_{m_{\quark\antiquark}^{\mathrm{cut}}}^{m_{\st}-m_{\gr}-m_{\tau}}
        dm_{\quark\antiquark}\,m_{\quark\antiquark}\,
        {d\Gamma(\stau_{\mathrm R}\to\tau\,\gravitino\,\quark\,\antiquark)
        \over dm_{\quark\antiquark}} 
\label{Eq:MtimesBranchingRatio}
\ee 
with
$\Gamma^{\rm{total}}_{\tilde{\tau}}\simeq\Gamma(\stau\to\tau\gravitino)$
and the invariant mass of the quark--antiquark pair
$m_{\quark\antiquark}$. We consider only quark--antiquark pairs with
$m_{\quark\antiquark}$ of at least
$m_{\quark\antiquark}^{\mathrm{cut}}=2~\GeV$,
i.e., the mass of a nucleon pair, since hadronic effects of mesons are
negligible for $\tau_{\st} \gtrsim
100~\seconds$~\cite{Kawasaki:2004qu}.

In our calculation of the 4-body decay
$\stau_{\mathrm R}\to\tau\,\gravitino\,\quark\,\antiquark$,
the exchange of virtual photons and virtual Z bosons is included. The
width of the Z~boson is taken into account by using the Breit--Wigner
form of the Z-boson propagator. We neglect the tau mass and assume
that the lightest neutralino is a pure bino with
$m_{\Bi}=1.1\,m_{\st}$ as already mentioned. The other neutralinos and
the squarks are assumed to be much heavier than the stau NLSP and the
bino. Depending on $m_{\quark\antiquark}$, up to five quark flavors
are considered.  The contributions from $\topquark\antitopquark$-final
states, which can appear for $m_{\quark\antiquark} \gtsim 350~\GeV$,
are subleading.

In Fig.~\ref{Fig:MassBoundsAHSNoCo} the thick solid (red) and thick
dashed (blue) curves show the BBN bounds from late hadronic and
electromagnetic energy injection, respectively. The regions inside or
to the right of the corresponding curves are excluded.
With~(\ref{Eq:YstauNoCo}), (\ref{Eq:MtimesBranchingRatio}), and
$\epsilon_{\EM}=0.5 E_{\tau}$, the curves in the left (right) plot are
obtained from the severe (conservative) upper limits on
$\xi_{\EM,\HAD}$ defined in Sec.~4.1 of Ref.~\cite{Steffen:2006hw}
based on results from Refs.~\cite{Kawasaki:2004qu,Cyburt:2002uv}.
Recall that we focus our investigation on late decays
$\tau_{\st}\gtrsim 100~\seconds$. For shorter lifetimes, additional
exclusion limits may arise from decays of the emitted $\tau$ into
mesons since these mesons can trigger proton--neutron interconversion
processes~\cite{Kawasaki:2004qu}.

An additional constraint from the observed Planck spectrum of the
cosmic microwave background (CMB) has been updated
recently~\cite{Lamon:2005jc}. The new CMB limit for
$\epsilon_{\EM}=0.5 E_{\tau}$ is expected to be similar to the thick
dashed (blue) line shown in the right plot of
Fig.~\ref{Fig:MassBoundsAHSNoCo}. Only in the `gap' region, we expect
the CMB limit to be more severe. The thick dashed (blue) line in the
left plot however will everywhere be more severe than the CMB limit.

Recently, it has been stressed that bound--state formation of
long-lived negatively charged particles with the primordial nuclei can
affect
BBN~\cite{Pospelov:2006sc,Kohri:2006cn,Kaplinghat:2006qr,Cyburt:2006uv}.
With the charged long-lived stau NLSP, these bound--state effects also
apply to the considered gravitino dark matter scenarios.  In
particular, a significant enhancement of $^6$Li production has been
found to imply the bound
$\tau_{\st}\ltsim 4\times 10^3~\seconds$~\cite{Pospelov:2006sc},
which excludes the $(\mgravitino,m_{\st})$ region below the thick
dotted (green) line shown in Fig.~\ref{Fig:MassBoundsAHSNoCo}.

The mass bounds shown in Fig.~\ref{Fig:MassBoundsAHSNoCo} are crucial
for gravitino dark matter studies, insights into the SUSY breaking
mechanism, and potential gravitino signatures at future colliders. The
appearance of a long-lived stau as the lightest Standard Model
superpartner in the collider detector will point to an extremely
weakly interacting LSP such as the gravitino or the
axino~\cite{Covi:2001nw,Brandenburg:2004du+X}. In the gravitino LSP
case, the cosmological constraints will provide an upper bound on
$\mgravitino$ and thereby an upper bound on the SUSY breaking scale
once $m_{\st}$ is measured. This $\mgravitino$ bound will imply upper
limits on $\Omega_{\gravitino}^{\NTP}$ and $T_{\Reheating}$.
While the microscopic measurement of $\MPl$ proposed
in~\cite{Buchmuller:2004rq,Martyn:2006as} could provide strong
evidence for the gravitino LSP, the required kinematical determination
of $\mgravitino$ appears to be feasible only for
$\mgravitino/m_{\st} \gtrsim 0.1$.
Unfortunately, this $(\mgravitino,m_{\st})$ region seems to be
excluded by the new $^6$Li
bounds~\cite{Pospelov:2006sc,Cyburt:2006uv}.
If these bounds cannot be avoided, alternative methods such as the
ones proposed in~\cite{Brandenburg:2005he,Steffen:2005cn} will become
essential to identify the gravitino as the LSP.

\end{document}